\documentclass[preprint,eqsecnum,aps,nofootinbib]{revtex4}
\usepackage[dvips]{graphicx}
\usepackage{amssymb}
\usepackage{amsmath}
\usepackage{amsmath, amsthm, amssymb, mathrsfs}
\usepackage{pstricks}

\newcommand{\METER}{\psframe( 0.00,-0.25)( 0.50, 0.25)%
                    \psarc[linewidth=0.2pt]( 0.25,-0.20){0.25}{50}{130}%
                    \psline[linewidth=0.2pt, arrows=->]( 0.25,-0.20)( 0.35, 0.2)
           }

\begin{document}

\title{Role of Three-Qubit Mixed-States Entanglement in Teleportation Scheme}
\author{
DaeKil Park}

\affiliation{Department of Physics, Kyungnam University, Masan,
631-701, Korea}

\author{Sayatnova Tamaryan}

\affiliation{Theory Department, Yerevan Physics Institute,
Yerevan-36, 375036, Armenia}

\author{Jin-Woo Son}

\affiliation{Department of Mathematics, Kyungnam University, Masan,
631-701, Korea}

\vspace{1.0cm}

\begin{abstract}
The bipartite quantum teleportation with a three-qubit mixture of GHZ and W states as 
a quantum channel is discussed. When the quantum channel is a mixed state consisting of
the GHZ state plus small perturbed W state, the entanglement of the channel becomes zero when
the average fidelity $\bar{F}_{GHZ}$ is less than $0.775$. This means that the mixed 
state cannot play an any role as a quantum channel at $\bar{F}_{GHZ} \leq 0.775$. For the 
case of the mixed state consisting of the W state plus small perturbed GHZ state it 
turn out that this mixed state cannot play a role as a quantum channel at 
$\bar{F}_W \leq 0.833$.
\end{abstract}


\maketitle

\section{Introduction}
Recently, there has been a flurry of activity in the entanglement of the quantum
states\cite{ami07}. It seems to be a genuine physical resource which makes quantum 
computer outperforms classical one\cite{vidal03-1}. 
It plays an important role in other branches of physics. For example, 
entanglement may give a 
promising approach to understand the information loss problem of the black hole
physics. The entanglement between matters and gravity\cite{terno05} may shed light on
the various difficulties arising from the black hole physics. Therefore, it is highly 
important to understand the general properties of the quantum entanglement in the 
context of quantum information theories\cite{nielsen00}.

The quantum entanglement for the two-qubit states is well understood regardless of 
pure or mixed states. For example, the entanglement of formation ${\cal E}$\cite{form1}
and the Groverian measure $G$\cite{biham01-1}, two of the basic entanglement measures,
for any two-qubit pure states can be straightly computed from the concurrence 
${\cal C}$ using a formula
\begin{equation}
\label{two1}
{\cal E}(\psi) = h \left( \frac{1 + \sqrt{1 - {\cal C}^2 (\psi)}}{2}\right)
\hspace{1.0cm}
G(\psi) = \frac{1}{\sqrt{2}} \left[1 - \sqrt{1 - {\cal C}^2 (\psi)} \right]^{1/2}
\end{equation}
where $h(x) \equiv -x \log_2 x - (1 - x) \log_2 (1 - x)$. For two-qubit state
$|\psi\rangle = \sum_{i,j=0}^1 a_{ij} |ij\rangle$ the concurrence ${\cal C}(\psi)$ becomes
\begin{equation}
\label{two2}
{\cal C}(\psi) = 2 |a_{00} a_{11} - a_{01} a_{10}|.
\end{equation}
Thus, it is maximal for Bell states and vanishes for factorized states. Combining 
(\ref{two1}) and (\ref{two2}), one can compute the entanglement of formation and the 
Groverian measure for all two-qubit pure states.

The entanglement for the mixed states is in general defined by making use of the 
convex roof construction\cite{benn96,uhlmann99-1}\footnote{For Groverian measure of mixed
states there is 
an entanglement monotone, which does not follow the convex roof construction. See
Ref.\cite{shapira06}}. For example, the concurrence for the two-qubit mixed state $\rho$
is defined as 
\begin{equation}
\label{two3}
{\cal C} (\rho) = \mbox{min} \sum_i p_{i} {\cal C} (\rho_i)
\end{equation}
where minimum is taken over all possible ensembles of pure states. The ensemble which gives 
the minimum value in Eq.(\ref{two3}) is called the optimal decomposition of the 
mixed state $\rho$. Few years ago W. K. Wootters\cite{form2,form3} has shown how to
construct the optimal ensembles for the arbitrary two-qubit mixed states by considering 
the time reversal operation of the spin-1/2 particles. Making use of these optimal 
decompositions, one can analytically compute the concurrence for all two-qubit states 
by a simple formula
\begin{equation}
\label{two4}
{\cal C} (\rho) = \mbox{max} (0, \lambda_1 - \lambda_2 - \lambda_3 - \lambda_4)
\end{equation}
where $\lambda_i$'s are the eigenvalues, in decreasing order, of the Hermitian matrix
$$
\sqrt{\sqrt{\rho} (\sigma_y \otimes \sigma_y) \rho^* (\sigma_y \otimes \sigma_y)
\sqrt{\rho}}. $$
Note that ${\cal E} (\psi)$ and $G (\psi)$ in Eq.(\ref{two1}) are 
monotonic functions with respect to ${\cal C} (\psi)$. This fact indicates that the 
optimal decompositions for the entanglement of formation and the Groverian measure are 
same with those for the concurrence. Thus one can compute ${\cal E}$, $G$ and ${\cal C}$
for all two-qubit states regardless of pure or mixed.

Recently, ${\cal E}$, $G$ and ${\cal C}$ for the various mixed states arising in the 
teleportation process through noisy channels have been explicitly computed\cite{08-mixed}.
Due to the noises the sender, Alice, cannot send the single-qubit state $|\psi_{in}\rangle$
to the receiver, Bob, perfectly. If Bob receives $\rho_{out}$, one can compute
$F \equiv \langle \psi_{in} | \rho_{out} |\psi_{in}\rangle$, which measures how well 
the teleportation job is performed. It is shown in Ref.\cite{08-mixed} that the mixed
states entanglements ${\cal E}$, $G$ and ${\cal C}$ all vanish when the average 
of $F$, say $\bar{F}$, is less than $2/3$, which corresponds to the best possible 
score when Alice and Bob communicate with each other through classical 
channel\cite{popescu94-1}. This fact implies that the mixed state entanglement is a genuine
physical resource for the teleportation process through noisy channels.

In this paper we would like to explore same issue with Ref.\cite{08-mixed} in the 
teleportation when the quantum channel is three-qubit mixed states. 
It is well-known that not only
two-qubit Bell state but also three-qubit GHZ\cite{karl98} and W\cite{agra06-1}
states
\begin{equation}
\label{ghzandw}
|\psi_{GHZ}\rangle = \frac{1}{\sqrt{2}} \left( |000\rangle + |111\rangle \right)
\hspace{1.0cm}
|\psi_{W}\rangle = \frac{1}{2} \left( |100\rangle + |010\rangle + \sqrt{2} 
                            |001\rangle \right)
\end{equation}
allow the perfect teleportation if Alice and Bob share the qubits appropriately at the 
initial stage..
The imperfect teleportation due to various noises was discussed in Ref.\cite{all1} when
GHZ and W states are prepared for two-party quantum teleportation through a noisy channel. 
The various three-qubit mixed states and the average fidelity $\bar{F}$ were explicitly
derived in the reference. It has been shown that the issue of robustness between GHZ and 
W, i.e. which state does lose less quantum information, in the noisy channels is completely
dependent on the type of noisy channel.

Then, it is natural to ask the following questions. Does the entanglement for the three-qubit
mixed state play an important role in the teleportation process as that for the 
two-qubit mixed state does as shown in Ref.\cite{08-mixed}? What is the relation between
$\bar{F}$ and entanglement? Do all entanglement measures vanish when $\bar{F} \leq 2/3$?
The purpose of this paper is to try to address these questions.

However, unfortunately, our knowledge on the entanglement of the three-qubit mixed state 
is extremely limited. Especially, the entanglement of formation and the concurrence are 
in fact defined in the bipartite system. Therefore, we cannot use them in the three-qubit
state\footnote{Recently, there was a try to extend the definition of the concurrence to
the multi-qubit system.\cite{concur-1,concur-2}}. Of course, the Groverian measure can be
defined in the arbitrary qubit states. Although, recently, the analytical computation
of the Groverian measure for several three-qubit pure states was 
done\cite{jung07-1,tama07-1,tama08-1,jung08-1}, still we do not know how to construct the 
optimal decomposition for the three-qubit mixed states to compute the Groverian measure.
Since, however, we are considering the two-party quantum teleportation with three-qubit
mixed state and the sender, Alice, has usually first two qubits, it is sufficient to 
consider the two-party concurrence ${\cal C}_{(AB)C}$ if $\rho^{ABC}$ is a given mixed 
state. The concurrence ${\cal C}_{(AB)C}$ can be computed for three-qubit pure states
by making use of the three-tangle\cite{tangle1}. Furthermore, although we can not compute 
it for arbitrary three-qubit mixed state, we know how to construct the optimal 
decomposition at least for the mixtures of GHZ and W states\cite{tangle2,tangle3}. Thus, 
we can use these mixture states to address the purpose of the present paper.

This paper is organized as follows. In section II we would like to briefly review the 
three-tangle and its optimal decomposition for the mixtures of GHZ and W states. In
this section we will compute ${\cal C}_{(AB)C}$ explicitly for the mixed state we will use 
in the teleportation process.
In section III we will compute the average fidelity $\bar{F}$ when the mixture consisting 
of the unperturbed GHZ state and small perturbed W state is used as a quantum channel.
It is shown that the entanglement of the quantum channel becomes zero at
$\bar{F} \leq 0.775$. This fact indicates that the bipartite teleportation with GHZ
state is more unstable than that with usual EPR state under the perturbed interaction.
In section IV we will compute the average fidelity $\bar{F}$ when the mixture consisting 
of the unperturbed W state and small perturbed GHZ state is used as a quantum channel.
unperturbed W and small perturbed GHZ states is used as a quantum channel.
It is shown that the entanglement of the quantum channel becomes zero at 
$\bar{F} \leq 0.833$.
In section V a brief discussion on the previous calculational results is given.

\section{Three-Tangle and Computation of ${\cal C}_{(AB)C}$}

\begin{figure}[ht!]
\begin{center}
\includegraphics[height=10.0cm]{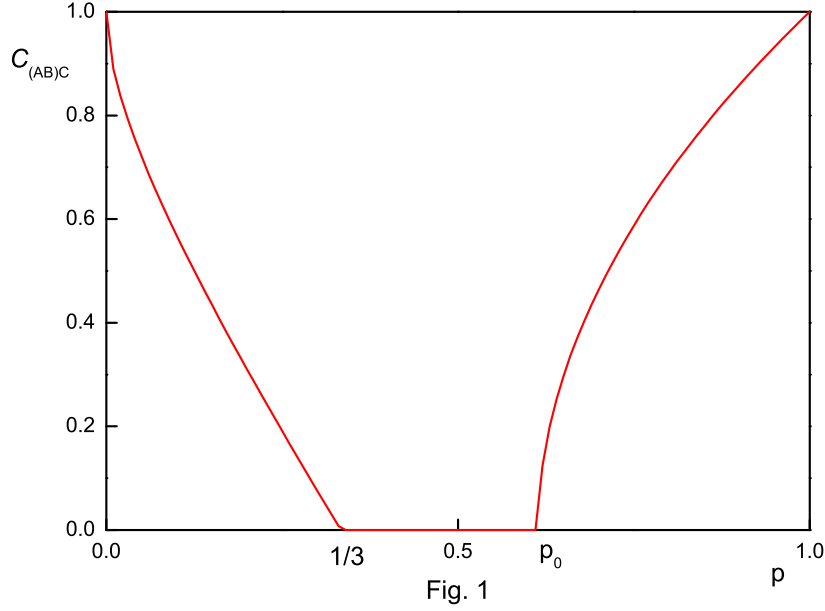}
\caption[fig1]{Plot of $p$-dependence of ${\cal C}_{(AB)C}$. The non-zero of 
${\cal C}_{(AB)C}$ at $p > p_0$ is mainly due to the three-tangle of the mixed state
$\rho^{QC}$ while the remaining non-vanishing value at $p < 1/3$ is due to the concurrences 
of the reduced states. The fact that ${\cal C}_{(AB)C} = 1$ at $p=0$ and $p=1$ implies that 
pure GHZ and pure W states are maximally entangled.}
\end{center}
\end{figure}

For three-qubit pure state $\rho^{ABC}$ the concurrences ${\cal C}_{AC}$ and 
${\cal C}_{BC}$ for the reduced states $\rho^{AC}$ and $\rho^{BC}$ satisfy the following
inequality\cite{tangle1}
\begin{equation}
\label{inequality1}
{\cal C}_{AC}^2 + {\cal C}_{BC}^2 \leq {\cal C}^2_{(AB)C}
\end{equation}
where ${\cal C}_{(AB)C}$ is a concurrence between a pair AB and the qubit C. The difference 
between rhs and lhs in Eq.(\ref{inequality1}) is defined as a three-tangle or residual 
entanglement:
\begin{equation}
\label{3-tangle}
\tau_{ABC} \equiv {\cal C}^2_{(AB)C} - \left({\cal C}_{AC}^2 + {\cal C}_{BC}^2 \right).
\end{equation}

For the state $|\psi\rangle = \sum_{i,j,k=0}^1 a_{ijk} |ijk\rangle$, $\tau_{ABC}$
becomes\cite{tangle1}
\begin{equation}
\label{3-tangle-1}
\tau_{ABC} = 4 |d_1 - 2 d_2 + 4 d_3|
\end{equation}
where
\begin{eqnarray}
\label{3-tangle-2}
& &d_1 = a^2_{000} a^2_{111} + a^2_{001} a^2_{110} + a^2_{010} a^2_{101} + 
                                                              a^2_{100} a^2_{011}
                                                              \\   \nonumber
& &d_2 = a_{000} a_{111} a_{011} a_{100} + a_{000} a_{111} a_{101} a_{010} + 
         a_{000} a_{111} a_{110} a_{001}
                                                              \\   \nonumber
& &\hspace{1.0cm} +
         a_{011} a_{100} a_{101} a_{010} + a_{011} a_{100} a_{110} a_{001} + 
         a_{101} a_{010} a_{110} a_{001}
                                                              \\   \nonumber
& &d_3 = a_{000} a_{110} a_{101} a_{011} + a_{111} a_{001} a_{010} a_{100}. 
\end{eqnarray}
Thus the generalized GHZ and W states defined 
\begin{equation}
\label{wg1}
|GHZ\rangle = a |100\rangle + b |111\rangle  \hspace{1.0cm}
|W\rangle = c |001\rangle + d |010\rangle + f |100\rangle 
\end{equation}
have
\begin{equation}
\label{3-tangle-3}
\tau_3^{GHZ} = 4 |a^2 b^2| \hspace{2.0cm} \tau_3^W = 0.
\end{equation}

For the mixed three-qubit state $\rho$ the three-tangle is defined by 
\begin{equation}
\label{optimal1}
\tau_3 (\rho) = \mbox{min} \sum_j p_j \tau_3 (\rho_j)
\end{equation}
where minimum is taken over all possible ensembles of pure states. Thus we should construct
the optimal decomposition to compute Eq.(\ref{optimal1}). The construction of the optimal
decompositions for the arbitrary three-qubit mixed states is an highly nontrivial and
formidable job and is yet unsolved. However, for the mixture of GHZ and W states
given by 
\begin{equation}
\label{mixture1}
\rho(p) = p |GHZ\rangle \langle GHZ| + (1-p) |W\rangle \langle W|
\end{equation}
the optimal decompositions has been constructed in Ref.\cite{tangle2,tangle3}.
The authors in these references used a fact that the state
\begin{equation}
\label{particu1}
|p,\varphi \rangle = \sqrt{p} |GHZ\rangle - \sqrt{1-p} e^{i \varphi} |W\rangle
\end{equation}
has nontrivial zero of the three-tangle at 
\begin{equation}
\label{zero1}
\varphi = n \frac{2 \pi}{3} \hspace{.5cm} (n \in \mathbb{N}) \hspace{2.0cm}
p = p_0 = \frac{s^{2/3}}{1 + s^{2/3}}
\end{equation}
where $s = 4 cdf/a^2 b$.

When $p \leq p_0$ one can derive an optimal ensemble in the form
\begin{equation}
\label{optimal2}
\rho(p) = \frac{p}{3 p_0} \left[ |p_0,0\rangle \langle p_0,0| + 
           |p_0,\frac{2\pi}{3}\rangle \langle p_0,\frac{2\pi}{3}| + 
          |p_0,\frac{4\pi}{3}\rangle \langle p_0,\frac{4\pi}{3}| \right] + 
         \left(1 - \frac{p}{p_0}\right) |W\rangle \langle W|
\end{equation}
which gives the vanishing three-tangle. The remaining optimal decompositions at 
$p > p_0$ have been derived in Ref.\cite{tangle2,tangle3} explicitly, and we do not 
want to repeat the derivation here. What we want to do here is to summarize the 
three-tangle as follows:
\begin{eqnarray}
\label{summary1}
\tau_3 (\rho(p)) = \left\{ \begin{array}{ll}
0 & \hspace{.5cm}   \mbox{for $0 \leq p \leq p_0$}  \\
\tau_3(p) & \hspace{.5cm}    \mbox{for $p_0 \leq p \leq p_1$}   \\
\tau_3^{conv} (p) & \hspace{.5cm}    \mbox{for $p_1 \leq p \leq 1$}
                            \end{array}               \right.
\end{eqnarray}
where
\begin{eqnarray}
\label{summary2}
& &\tau_3(p) = \tau_3^{GHZ} |p^2 - \sqrt{p (1 - p)^3} s|   \\   \nonumber
& &\tau_3^{conv} (p) = \tau_3^{GHZ} \left[ \frac{p - p_1}{1 - p_1} + \frac{1 - p}{1 - p_1}
\left(p_1^2 - \sqrt{p_1 (1 - p_1)^2} s \right) \right]
\end{eqnarray}
and
\begin{equation}
\label{summary3}
p_0 = \frac{s^{2/3}}{1 + s^{2/3}} \hspace{1.0cm}
p_1 = \mbox{max} \left(p_0, \frac{1}{2} + \frac{1}{2\sqrt{1 + s^2}} \right).
\end{equation}

As we mentioned in the previous section we will consider in this paper the quantum 
teleportation with a quantum channel
\begin{equation}
\label{mixture2}
\rho^{QC} = p |\psi_{GHZ}\rangle \langle \psi_{GHZ}| + (1-p) |\psi_W\rangle \langle \psi_W|
\end{equation}
where $|\psi_{GHZ}\rangle$ and $|\psi_W\rangle$ are given in Eq.(\ref{ghzandw}). Comparing
Eq.(\ref{mixture1}) with Eq.(\ref{mixture2}), we have $a=b=c=1/\sqrt{2}$ and $d=f=1/2$, 
which give
\begin{equation}
\label{consts}
s=2, \hspace{.5cm} \tau_3^{GHZ} = 1, \hspace{.5cm}
p_0 = \frac{2^{2/3}}{1 + 2^{2/3}} \sim 0.614, \hspace{.5cm}
p_1 = \frac{1 + \sqrt{5}}{2 \sqrt{5}} \sim 0.724.
\end{equation}
Thus the three-tangle for $\rho^{QC}$ becomes
\begin{eqnarray}
\label{residual1}
\tau_3 (\rho^{QC}) = \left\{    \begin{array}{ll}
0 & \hspace{.5cm}   \mbox{for $0 \leq p \leq p_0$}  \\
|p^2 - 2 \sqrt{p (1-p)^3}| & \hspace{.5cm}    \mbox{for $p_0 \leq p \leq p_1$}   \\
\frac{1}{1-p_1} \left[(1 - t_1) p - (p_1 - t_1) \right] 
                                   & \hspace{.5cm}    \mbox{for $p_1 \leq p \leq 1$}
                             \end{array}               \right.
\end{eqnarray}
where $t_1 \equiv p_1^2 - 2 \sqrt{p_1 (1-p_1)^3} \sim 0.276$.

Since it is simple to derive the reduced states from $\rho^{QC}$, one can easily compute
the concurrences ${\cal C}_{AB}$, ${\cal C}_{AC}$ and ${\cal C}_{BC}$ following
Wootters procedure\cite{form2,form3}, whose explicit expressions are 
\begin{eqnarray}
\label{reduced-conc}
& &{\cal C}_{AB} = \left\{      \begin{array}{ll}
\frac{1 - p - 2 \sqrt{p}}{2}  &   \hspace{.5cm} \mbox{for $0 \leq p \leq 3 - 2 \sqrt{2}$} \\
0  &  \hspace{.5cm} \mbox{for $3 - 2 \sqrt{2} \leq p \leq 1$}
                           \end{array}               \right.
                                                              \\   \nonumber
& &{\cal C}_{AC} = {\cal C}_{BC} = \left\{         \begin{array}{ll}
\frac{1}{\sqrt{2}} \left[ (1 - p) - \sqrt{p (1 + p)} \right]  &  \hspace{.5cm}
\mbox{for $0 \leq p \leq \frac{1}{3}$}                               \\
0   &  \hspace{.5cm} \mbox{for $\frac{1}{3} \leq p \leq 1$}
                          \end{array}               \right.
\end{eqnarray}
Thus one can compute ${\cal C}_{(AB)C}$ as
\begin{equation}
\label{final-ent}
{\cal C}_{(AB)C} = \sqrt{ {\cal C}^2_{AC} + {\cal C}^2_{BC} + \tau_3 (\rho^{QC})}.
\end{equation}

The $p$-dependence of ${\cal C}_{(AB)C}$ is plotted in Fig. 1. This figure shows that 
${\cal C}_{(AB)C} = 1$ at $p=0$ and $p=1$, which indicates that pure GHZ and pure W 
states are maximally entangled. This fact also indicates that the two-party teleportation
with $p=0$ and $p=1$ states will be perfect. This will be confirmed in next section by
showing that the average fidelity $\bar{F}$ becomes unit at these points. The non-vanishing
${\cal C}_{(AB)C}$ at $p < 1/3$ is due to the concurrences of the reduced states while
the remaining non-vanishing value at $p > p_0$ is due to the three-tangle. This fact 
indicates that the entanglement of pure GHZ states mainly comes from the three-tangle while
that of pure W state comes from the entanglement of its reduced states.

\section{Quantum Teleportation with large $p$ state}

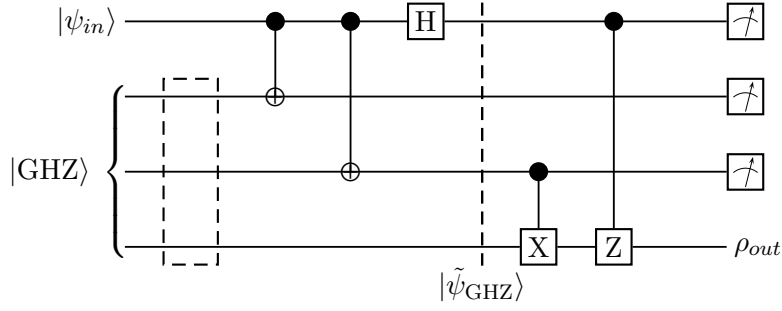
\begin{figure}
\begin{center}
\begin{pspicture}(-5,1.5)(5,6)

\psframe(-0.25, 5.25)(0.25, 5.75) \rput[c](0.00, 5.5){H}
\psframe( 1.25, 2.25)(1.75, 2.75) \rput[c](1.5, 2.5){X}
\psframe( 2.25, 2.25)(2.75, 2.75) \rput[c](2.5, 2.5){Z}
\psframe[linestyle=dashed](-3.5, 2.25)(-2.75, 4.75)

\psline[linewidth=0.7pt](-4.0, 5.5)(-0.25, 5.5) 
\psline[linewidth=0.7pt](0.25, 5.5)(4.0, 5.5)
\psline[linewidth=0.7pt](-4.0, 4.5)(4.0, 4.5)
\psline[linewidth=0.7pt](-4.0, 3.5)(4.0, 3.5)
\psline[linewidth=0.7pt](-4.0, 2.5)(1.25, 2.5)
\psline[linewidth=0.7pt](1.75, 2.5)(2.25, 2.5)
\psline[linewidth=0.7pt](2.75, 2.5)(4.0, 2.5)

\psline[linewidth=0.7pt](-2.0, 5.5)(-2.0, 4.5)
\psline[linewidth=0.7pt](-1.0, 5.5)(-1.0, 3.5)
\psline[linewidth=0.7pt](1.5, 3.5)(1.5, 2.75)
\psline[linewidth=0.7pt](2.5, 5.5)(2.5, 2.75)

\psline[linestyle=dashed, linewidth=0.9pt]( 0.75, 5.75)( 0.75, 2.25)

\psdot[dotstyle=oplus, dotscale=2](-2.0, 4.5)
\psdot[dotstyle=*, dotscale=2](-2.0, 5.5)
\psdot[dotstyle=oplus, dotscale=2](-1.0, 3.5)
\psdot[dotstyle=*, dotscale=2](-1.0, 5.5)
\psdot[dotstyle=*, dotscale=2]( 1.5, 3.5)
\psdot[dotstyle=*, dotscale=2]( 2.5, 5.5)

\rput[c](-4.0, 3.5){$\left\{ \begin{array}{r} \\ \\ \\ \end{array}  \right.$}

\rput[r](-4.1, 5.5){$\lvert \psi_{in} \rangle$}
\rput[c](-5.0, 3.5){$\lvert \textrm{GHZ} \rangle$}
\rput[l]( 4.1, 2.5){$ \rho_{out} $}
\rput[c]( 0.75, 2.0){$\lvert \tilde{\psi}_{\textrm{GHZ}} \rangle$}

\rput[c]( 4.0, 5.5){\METER}
\rput[c]( 4.0, 4.5){\METER}
\rput[c]( 4.0, 3.5){\METER}

\end{pspicture}
\end{center}
\caption{A quantum circuit for quantum teleportation through noisy channels with GHZ state. 
The top three lines belong to Alice while the bottom line belongs to Bob.
The dotted box represents small perturbation, which makes the quantum channel to be mixed
state.}
\end{figure}

In this section we consider the quantum teleportation with a mixed state 
$\rho^{QC}$ given in Eq.(\ref{mixture2}) when $p$ is large. The state $\rho^{QC}$ with 
large $p$ can be regarded as a mixed state which consists of the GHZ state plus small
perturbed W state. Therefore, we use a teleportation scheme with GHZ state, whose 
quantum circuit is given in Fig. 2.

Now, we assume that the sender, called Alice, who has first two qubits in $\rho^{QC}$, wants
to send a single qubit
\begin{equation}
\label{in-1}
|\psi_{in}\rangle = \cos \left(\frac{\theta}{2}\right) e^{i \phi/2} |0\rangle + 
                    \sin \left(\frac{\theta}{2}\right) e^{-i \phi/2} |1\rangle
\end{equation}
to the receiver, called Bob, who has last qubit in $\rho^{QC}$. Then Fig. 2 implies that
the state $\rho_{out}$, which Bob has finally, becomes
\begin{equation}
\label{out-1}
\rho_{out} = \mbox{Tr}_{1,2,3} \left[U_{GHZ} \left(\rho_{in} \otimes \rho^{QC}\right) 
U^{\dagger}_{GHZ}
                                                            \right]
\end{equation}
where $\mbox{Tr}_{1,2,3}$ is partial trace over Alice's qubits and 
$\rho_{in} = |\psi_{in}\rangle \langle \psi_{in}|$. The unitary operator $U_{GHZ}$ can be 
read directly from Fig. 2 and its explicit expression is 
\begin{eqnarray}
\label{Ughz}
U_{GHZ} = \frac{1}{\sqrt{2}} \left(   \begin{array}{cccccccccccccccc}
1 & 0 & 0 & 0 & 0 & 0 & 0 & 0 & 0 & 0 & 0 & 0 & 0 & 0 & 1 & 0  \\
0 & 1 & 0 & 0 & 0 & 0 & 0 & 0 & 0 & 0 & 0 & 0 & 0 & 0 & 0 & 1  \\
0 & 0 & 0 & 1 & 0 & 0 & 0 & 0 & 0 & 0 & 0 & 0 & 0 & 1 & 0 & 0  \\
0 & 0 & 1 & 0 & 0 & 0 & 0 & 0 & 0 & 0 & 0 & 0 & 1 & 0 & 0 & 0  \\
0 & 0 & 0 & 0 & 1 & 0 & 0 & 0 & 0 & 0 & 1 & 0 & 0 & 0 & 0 & 0  \\
0 & 0 & 0 & 0 & 0 & 1 & 0 & 0 & 0 & 0 & 0 & 1 & 0 & 0 & 0 & 0  \\
0 & 0 & 0 & 0 & 0 & 0 & 0 & 1 & 0 & 1 & 0 & 0 & 0 & 0 & 0 & 0  \\
0 & 0 & 0 & 0 & 0 & 0 & 1 & 0 & 1 & 0 & 0 & 0 & 0 & 0 & 0 & 0  \\
1 & 0 & 0 & 0 & 0 & 0 & 0 & 0 & 0 & 0 & 0 & 0 & 0 & 0 & -1 & 0  \\
0 & -1 & 0 & 0 & 0 & 0 & 0 & 0 & 0 & 0 & 0 & 0 & 0 & 0 & 0 & 1  \\
0 & 0 & 0 & 1 & 0 & 0 & 0 & 0 & 0 & 0 & 0 & 0 & 0 & -1 & 0 & 0  \\
0 & 0 & -1 & 0 & 0 & 0 & 0 & 0 & 0 & 0 & 0 & 0 & 1 & 0 & 0 & 0  \\
0 & 0 & 0 & 0 & 1 & 0 & 0 & 0 & 0 & 0 & -1 & 0 & 0 & 0 & 0 & 0  \\
0 & 0 & 0 & 0 & 0 & -1 & 0 & 0 & 0 & 0 & 0 & 1 & 0 & 0 & 0 & 0  \\
0 & 0 & 0 & 0 & 0 & 0 & 0 & 1 & 0 & -1 & 0 & 0 & 0 & 0 & 0 & 0  \\
0 & 0 & 0 & 0 & 0 & 0 & -1 & 0 & 1 & 0 & 0 & 0 & 0 & 0 & 0 & 0  
\end{array}                      \right).
\end{eqnarray}
Inserting Eq.(\ref{mixture2}), Eq.(\ref{in-1}), and Eq.(\ref{Ughz}) into Eq.(\ref{out-1}),
one can compute $\rho_{out}$ straightforwardly. 

In order to quantify how much information is preserved or lost during teleportation we 
consider a quantity
\begin{equation}
\label{fidel-1}
F(\theta, \phi) = \langle \psi_{in} | \rho_{out} | \psi_{in} \rangle,
\end{equation}
which is the square of the usual fidelity defined 
$F(\rho, \sigma) = \mbox{Tr} \sqrt{\rho^{1/2} \sigma \rho^{1/2}}$. Thus, $F=1$ implies the 
perfect teleportation.

For our case $F_{GHZ} (\theta, \phi)$ becomes
\begin{equation}
\label{ghz-fidel-1}
F_{GHZ} (\theta, \phi) = \frac{1}{8} \left[ (3 + 5 p) - (1 - p) \cos (2 \theta) \right].
\end{equation}
When $p=1$, $F_{GHZ}$ becomes one. This means that the pure GHZ state allows the perfect
teleportation. 

Now we define the average fidelity in a form
\begin{equation}
\label{fidel-2}
\bar{F} \equiv \frac{1}{4 \pi} \int_0^{2\pi} d\phi \int_0^{\pi} d\theta \sin \theta
             F(\theta, \phi).
\end{equation}
For our case $\bar{F}_{GHZ}$ becomes
\begin{equation}
\label{ghz-fidel-2}
\bar{F}_{GHZ} = \frac{5 + 7 p}{12}.
\end{equation}
When $p=1$, $\bar{F}_{GHZ}$ becomes one again.

\section{Quantum teleportation with small $p$ state}

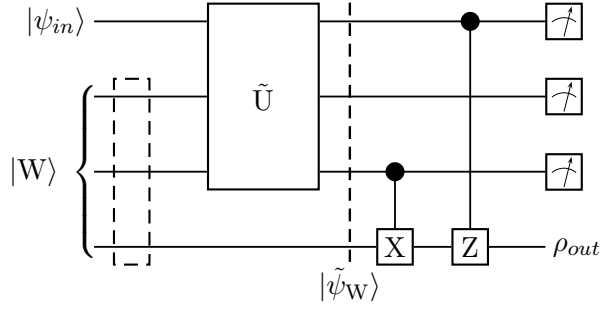
\begin{figure}
\begin{center}
\begin{pspicture}(-4,2.0)(4,6)
\psframe(-1.5, 3.25)( 0.00, 5.75) \rput[c](-0.75, 4.5){$\tilde{\textrm{U}}$}
\psframe( 0.75, 2.25)( 1.25, 2.75)\rput[c]( 1.00, 2.50){X}
\psframe( 1.75, 2.25)( 2.25, 2.75)\rput[c]( 2.00, 2.50){Z}
\psframe[linestyle=dashed](-2.75, 4.75)(-2.25, 2.25)

\rput[c]( 3.0, 5.50){\METER}
\rput[c]( 3.0, 4.50){\METER}
\rput[c]( 3.0, 3.50){\METER}

\psline[linewidth=0.7pt](-3.0, 5.5)(-1.50, 5.5)
\psline[linewidth=0.7pt]( 0.0, 5.5)( 3.00, 5.5)

\psline[linewidth=0.7pt](-3.0, 4.5)(-1.50, 4.5)
\psline[linewidth=0.7pt]( 0.0, 4.5)( 3.00, 4.5)

\psline[linewidth=0.7pt](-3.0, 3.5)(-1.50, 3.5)
\psline[linewidth=0.7pt]( 0.0, 3.5)( 3.00, 3.5)

\psline[linewidth=0.7pt](-3.0, 2.5)( 0.75, 2.5)
\psline[linewidth=0.7pt]( 1.25, 2.5)( 1.75, 2.5)
\psline[linewidth=0.7pt]( 2.25, 2.5)( 3.00, 2.5)

\psline[linestyle=dashed, linewidth=0.9pt]( 0.40, 5.75)( 0.40, 2.30)

\psline[linewidth=0.7pt]( 1.00, 3.50)( 1.00, 2.75)
\psline[linewidth=0.7pt]( 2.00, 5.50)( 2.00, 2.75)

\psdot[dotstyle=*, dotscale=2]( 1.00, 3.50)
\psdot[dotstyle=*, dotscale=2]( 2.00, 5.50)

\rput[r](-3.1, 5.5){$\lvert \psi_{in} \rangle$}
\rput[l]( 3.1, 2.5){$ \rho_{out} $}
\rput[c](-3.8, 3.5){$\lvert \textrm{W} \rangle$}
\rput[c](-3.0, 3.5){$\left\{ \begin{array}{r} \\ \\ \\ \end{array}  \right.$}

\rput[c](0.40, 2.00){$\lvert \tilde{\psi}_{\textrm{W}} \rangle$}
\end{pspicture}
\end{center}
\caption{A quantum circuit for quantum teleportation through noisy channels with W state. 
The top three lines belong to Alice while the bottom line belongs to Bob.
The dotted box represents small perturbation, which makes the quantum channel to be mixed
state. The unitary operator $\tilde{U}$ makes $|\tilde{\psi}_W\rangle$ coincide with
$|\tilde{\psi}_{GHZ}\rangle$ expressed in Fig. 2}
\end{figure}


Since the state $\rho^{QC}$ in Eq.(\ref{mixture2}) with small $p$ can be regarded as a 
mixed state which consists of the W state plus small perturbed GHZ state, we use a 
teleportation scheme with a pure W states. The quantum circuit for this scheme is 
given in Fig.3. In this figure the unitary operator $\tilde{U}$ is introduced to make
$|\tilde{\psi}_W\rangle$ to be same with $|\tilde{\psi}_{GHZ}\rangle$ in Fig. 2. The 
explicit expression of $\tilde{U}$ is given in Eq.(3.1) of Ref.\cite{all1}.

The final state $\rho_{out}$ which Bob has finally becomes
\begin{equation}
\label{out-2}
\rho_{out} = \mbox{Tr}_{1,2,3} \left[ U_W \left(\rho_{in} \otimes \rho^{QC} \right)
                                      U_W^{\dagger}   \right]
\end{equation}
where $\mbox{Tr}_{1,2,3}$ is partial trace over Alice's qubits and 
$\rho_{in} = |\psi_{in}\rangle \langle \psi_{in}|$. The explicit expression of $U_W$ can
be read directly from Fig. 3 in the form:
\begin{eqnarray}
\label{Uw}
U_W = \frac{1}{2} \left(             \begin{array}{cccccccccccccccc}
0 & 0 & 1 & 0 & 1 & 0 & 0 & 0 & \sqrt{2} & 0 & 0 & 0 & 0 & 0 & 0 & 0    \\
0 & 0 & 0 & 1 & 0 & 1 & 0 & 0 & 0 & \sqrt{2} & 0 & 0 & 0 & 0 & 0 & 0    \\
0 & 0 & 0 & 0 & 0 & 0 & 0 & 2 & 0 & 0 & 0 & 0 & 0 & 0 & 0 & 0    \\
0 & 0 & 0 & 0 & 0 & 0 & 2 & 0 & 0 & 0 & 0 & 0 & 0 & 0 & 0 & 0    \\
0 & 0 & 0 & 0 & 0 & 0 & 0 & 0 & 0 & 0 & 0 & 0 & 0 & 0 & 2 & 0    \\
0 & 0 & 0 & 0 & 0 & 0 & 0 & 0 & 0 & 0 & 0 & 0 & 0 & 0 & 0 & 2    \\
0 & \sqrt{2} & 0 & 0 & 0 & 0 & 0 & 0 & 0 & 0 & 0 & 1 & 0 & 1 & 0 & 0    \\
\sqrt{2} & 0 & 0 & 0 & 0 & 0 & 0 & 0 & 0 & 0 & 1 & 0 & 1 & 0 & 0 & 0    \\
0 & 0 & 1 & 0 & 1 & 0 & 0 & 0 & -\sqrt{2} & 0 & 0 & 0 & 0 & 0 & 0 & 0    \\
0 & 0 & 0 & -1 & 0 & -1 & 0 & 0 & 0 & \sqrt{2} & 0 & 0 & 0 & 0 & 0 & 0    \\
0 & 0 & 0 & \sqrt{2} & 0 & -\sqrt{2} & 0 & 0 & 0 & 0 & 0 & 0 & 0 & 0 & 0 & 0    \\
0 & 0 & -\sqrt{2} & 0 & \sqrt{2} & 0 & 0 & 0 & 0 & 0 & 0 & 0 & 0 & 0 & 0 & 0    \\
0 & 0 & 0 & 0 & 0 & 0 & 0 & 0 & 0 & 0 & \sqrt{2} & 0 & -\sqrt{2} & 0 & 0 & 0    \\
0 & 0 & 0 & 0 & 0 & 0 & 0 & 0 & 0 & 0 & 0 & -\sqrt{2} & 0 & \sqrt{2} & 0 & 0    \\
0 & \sqrt{2} & 0 & 0 & 0 & 0 & 0 & 0 & 0 & 0 & 0 & -1 & 0 & -1 & 0 & 0    \\
-\sqrt{2} & 0 & 0 & 0 & 0 & 0 & 0 & 0 & 0 & 0 & 1 & 0 & 1 & 0 & 0 & 0    \\
                                      \end{array}         \right).
\end{eqnarray}

Then $F_W(\theta, \phi)$ and $\bar{F}_W$ can be straightforwardly computed. It is shown 
that $F_W(\theta, \phi)$ is independent of the angle parameters. Thus $F_W(\theta, \phi)$
is same with the average fidelity in a form:
\begin{equation}
\label{w-fidel-1}
\bar{F}_W = F_W(\theta, \phi) = 1 - \frac{p}{2}.
\end{equation}
When $p=0$, $\bar{F}_W$ becomes one, which indicates that the pure W state allows the 
perfect teleportation.

\section{Discussion}

\begin{figure}[ht!]
\begin{center}
\includegraphics[height=10.0cm]{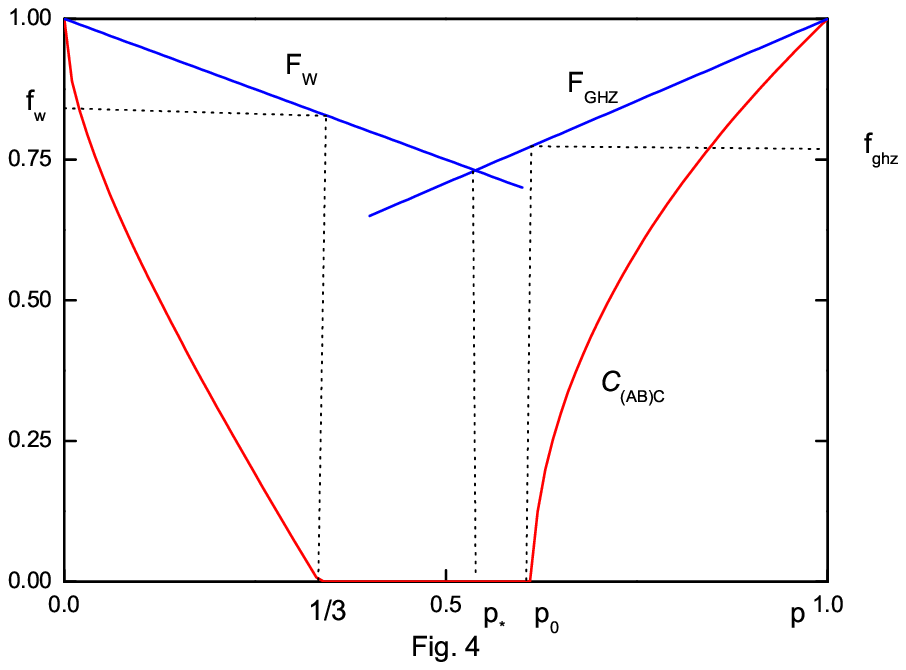}
\caption[fig1]{Plot of $p$-dependence of $\bar{F}_{GHZ}$ and $\bar{F}_W$. For comparison
we plot the $p$-dependence of ${\cal C}_{(AB)C}$ together. For the case of GHZ with small
perturbed W state the entanglement becomes zero at $\bar{F}_{GHZ} < f_{ghz} \sim 0.775$.
For the case of W with small perturbed GHZ state ${\cal C}_{(AB)C}$ becomes zero at 
$\bar{F}_W < f_w \sim 0.833$. In this aspect we can say that GHZ state is more robust 
in the teleportation process than W state under the perturbed interaction.}
\end{center}
\end{figure}

Fig. 4 is a plot of $p$-dependence of $\bar{F}_{GHZ}$ and $\bar{F}_W$ together with 
${\cal C}_{(AB)C}$. As expected, $\bar{F}_{GHZ}$ and $\bar{F}_W$ become unit when 
$p=1$ and $p=0$, respectively. At $p=p_0$ $\bar{F}_{GHZ}$ becomes $f_{ghz}$, where
\begin{equation}
\label{fghz}
f_{ghz} = \frac{1}{12} (5 + p_0) \sim 0.774549.
\end{equation}
When $\bar{F}_{GHZ} \leq f_{ghz}$, ${\cal C}_{(AB)C}$ of $\rho^{QC}$ with large $p$ becomes
zero. This means that the mixed state consisting of the GHZ plus small perturbed W state
cannot play a role as a quantum channel in the teleportation process when 
$\bar{F}_{GHZ} \leq f_{ghz}$. In this sense $f_{ghz}$ is something like the critical
value in the average fidelity $\bar{F}_{GHZ}$. It is interesting to note 
$f_{ghz} > 2/3$, where $\bar{F} = 2/3$ plays same role in the bipartite teleportation 
through noisy channel when EPR state is used as a quantum channel\cite{08-mixed}. This fact
indicates that the bipartite teleportation with GHZ state is much more unstable than
teleportation with EPR state under the perturbed interaction.

The critical value of $\bar{F}_W$ is $f_w$, where
\begin{equation}
\label{fw}
f_w = \frac{5}{6} \sim 0.833333.
\end{equation}
Since $f_w > f_{ghz}$, the bipartite teleportation with GHZ state is more stable than that
with W state. In this sense we can say that GHZ state is more robust than W state under
the perturbation.

There is another point we would like to consider. Fig. 4 shows that 
$\bar{F}_W \geq \bar{F}_{GHZ}$ in the region $0 \leq p \leq p_{*} = 7/13 \sim 0.538$. Since
$p_* > 1/2$, this means that $\rho^{QC}$ can be regarded as a W state with perturbed GHZ
state in the more wide range of $p$. In this aspect we can say that W state is more 
robust than GHZ state in the perturbed interaction.

Although we have considered the bipartite teleportation with a three-qubit mixed state
$\rho^{QC}$, this is not a teleportation through noisy channels because $\rho^{QC}$ is not
derived from the master equation\cite{lind76}. Many three-qubit mixed states are 
explicitly derived in Ref.\cite{all1} in the teleportation through noisy channels. For
example, the mixed state derived from W state with ($L_{2,x}$, $L_{3,x}$, $L_{4,x}$) 
noisy channel is 
\begin{eqnarray}
\label{qo-w2}
\varepsilon_x (\rho_W) = \frac{1}{16}
\left(           \begin{array}{cccccccc}
2 \alpha_2 & 0 & 0 & \sqrt{2} \alpha_2 & 0 & \sqrt{2} \alpha_2 & \alpha_2 & 0  \\
0 & 2 \alpha_1 & \sqrt{2} \alpha_1 & 0 & \sqrt{2} \alpha_1 & 0 & 0 & \alpha_3  \\
0 & \sqrt{2} \alpha_1 & 2 \beta_+ & 0 & \alpha_1 & 0 & 0 & \sqrt{2} \alpha_3  \\\sqrt{2} \alpha_2 & 0 & 0 & 2 \beta_- & 0 & \alpha_4 & \sqrt{2} \alpha_4 & 0   \\
0 & \sqrt{2} \alpha_1 & \alpha_1 & 0 & 2 \beta_+ & 0 & 0 & \sqrt{2} \alpha_3   \\
\sqrt{2} \alpha_2 & 0 & 0 & \alpha_4 & 0 & 2 \beta_- & \sqrt{2} \alpha_4 & 0   \\
\alpha_2 & 0 & 0 & \sqrt{2} \alpha_4 & 0 & \sqrt{2} \alpha_4 & 2 \alpha_4 & 0  \\
0 & \alpha_3 & \sqrt{2} \alpha_3 & 0 & \sqrt{2} \alpha_3 & 0 & 0 & 2 \alpha_3
                 \end{array}
                                                            \right)
\end{eqnarray}
where
\begin{eqnarray}
\label{def-w1}
& &\alpha_1 = 1 + e^{-2 \kappa t} + e^{-4 \kappa t} + e^{-6 \kappa t}  \\  \nonumber
& &\alpha_2 = 1 + e^{-2 \kappa t} - e^{-4 \kappa t} - e^{-6 \kappa t}  \\  \nonumber
& &\alpha_3 = 1 - e^{-2 \kappa t} - e^{-4 \kappa t} + e^{-6 \kappa t}  \\  \nonumber
& &\alpha_4 = 1 - e^{-2 \kappa t} + e^{-4 \kappa t} - e^{-6 \kappa t}  \\  \nonumber
& &\beta_{\pm} = 1 \pm e^{-6 \kappa t}.
\end{eqnarray}
Unfortunately, still there is no method, as far as we know, for the construction of the 
optimal decomposition of $\varepsilon_x (\rho_W)$. Thus it is highly nontrivial and 
formidable task to compute the three-tangle for this mixed state. We would like to 
develop a computational technique, which enables us to compute the three-tangle for 
arbitrary three-qubit mixed states or, at least, the various mixed states derived 
from master equation in Ref.\cite{all1}. If then, we may understand more deeply the role
of entanglement in the various quantum algorithms. 

{\bf Acknowledgement}: 
This work was supported by the Kyungnam University
Foundation Grant, 2008.

\end{document}